\documentclass[aps,prb,twocolumn,floatfix,preprintnumbers,amsmath,amssymb,superscriptaddress]{revtex4-1}

\usepackage{graphicx}% Include figure files
\usepackage{graphics}
\usepackage{dcolumn}% Align table columns on decimal point. Column format d.
\usepackage{bm}% bold math

\begin{document}
\DeclareGraphicsExtensions{.pdf,.png,.gif,.jpg}

\title{Excitations in a quantum spin liquid with random bonds}

 \author{D. H\"uvonen}
 \affiliation{Neutron Scattering and Magnetism, Laboratory for Solid State Physics, ETH Zurich, Switzerland.}

 \author{S. Zhao}
\altaffiliation{Current address: FBS Swiss Federal Institute of Technology (EPFL), CH-1015 Lausanne, Switzerland}
 \affiliation{Neutron Scattering and Magnetism, Laboratory for Solid State Physics, ETH Zurich, Switzerland.}

% \author{T. Yankova}
% \affiliation{Neutron Scattering and Magnetism, Laboratory for Solid State Physics, ETH Zurich, Switzerland.}
% \affiliation{Chemistry Dept., Lomonosov Moscow State University, Russia.}
 \author{G. Ehlers}
 \affiliation{Oak Ridge National Laboratory, Oak Ridge, TN, USA}
%\author{C. Niedermayer}
% \affiliation{Laboratory for Neutron Scattering, Paul Scherrer Institute, CH-5232 Villigen, Switzerland.}
% \author{M. Laver}
% \affiliation{Laboratory for Neutron Scattering, Paul Scherrer Institute, CH-5232 Villigen, Switzerland.}
% \affiliation{Materials Research Division, Risø DTU, Technical University of Denmark, DK-4000 Roskilde, Denmark.}
% \affiliation{Nano Science Center, Niels Bohr Institute, University of Copenhagen, DK-2100 Copenhagen, Denmark.}
 \author{M. M\aa nsson}
 \affiliation{Neutron Scattering and Magnetism, Laboratory for Solid State Physics, ETH Zurich, Switzerland.}
 \author{S.N. Gvasaliya}
 \affiliation{Neutron Scattering and Magnetism, Laboratory for Solid State Physics, ETH Zurich, Switzerland.}
 \author{A. Zheludev}
 \homepage{http://www.neutron.ethz.ch/}
 \affiliation{Neutron Scattering and Magnetism, Laboratory for Solid State Physics, ETH Zurich, Switzerland.}
\date{\today}

\begin{abstract}
We present results of inelastic neutron scattering study on two bond disordered quasi two-dimensional quantum magnets (C$_4$H$_{12}$N$_2$)Cu$_2$(Cl$_{1-x}$Br$_x$)$_6$ with x=0.035 and 0.075.
We observe the increase of spin gap, reduction of magnon bandwidth and a decrease of magnon lifetimes compared to x=0 sample.
Additional magnon damping is observed at higher energies away from zone center which is found to follow the density of single particle states.

\end{abstract}


\maketitle

\section{Introduction}
\begin{figure}[tb]
\includegraphics[width=8.5cm]{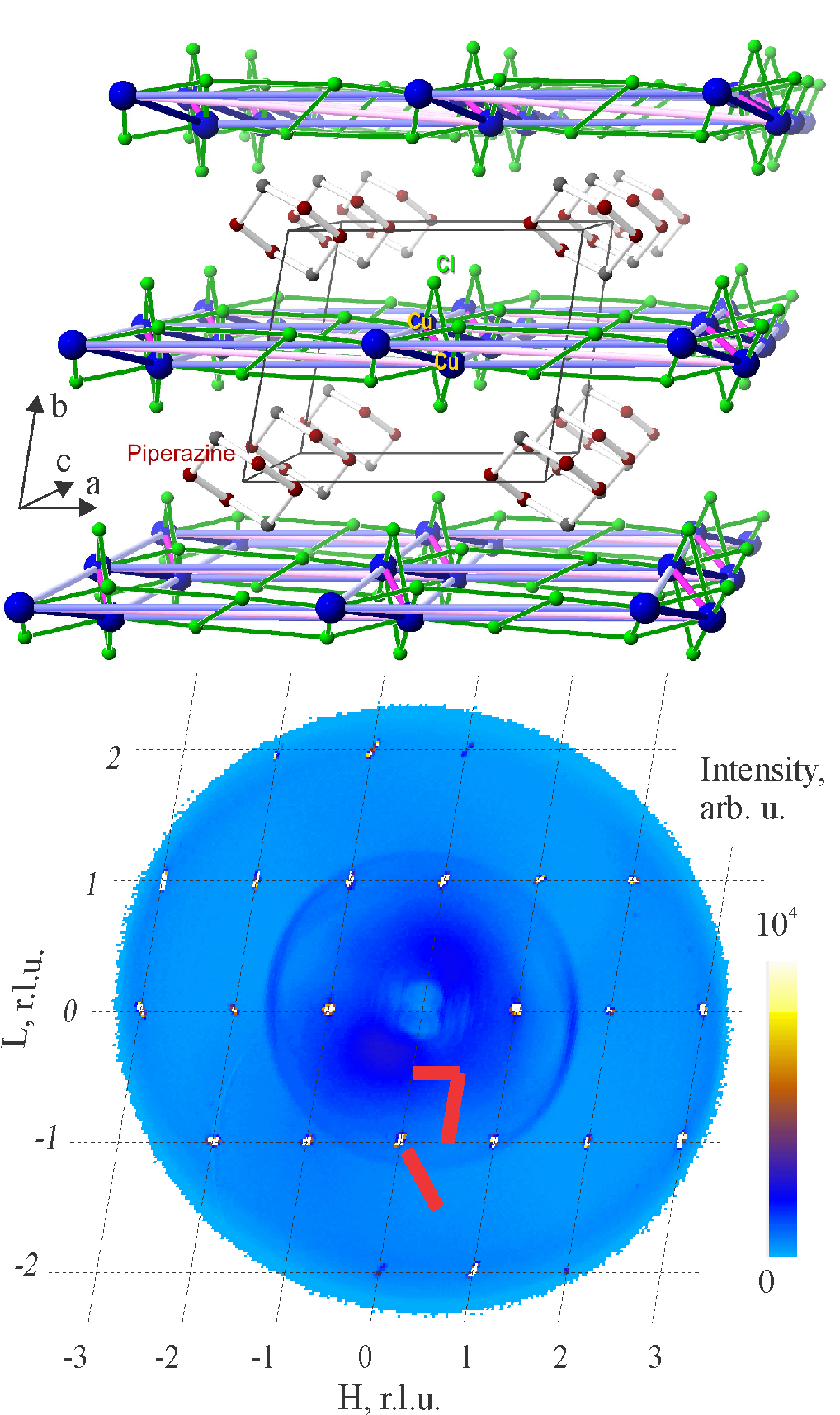}\\
\caption{(Color online) \emph{Top:} Structure of PHCC, Magnetic Cu$^{2+}$ (blue) ions are coupled via Cl (green) ions to form 2D layers separated by piperazine molecules. Thick lines within layers indicate possible superexchange pathways. \emph{Bottom:} Elastic neutron scattering intensity in the (H,L) plane (integrated over K direction) for 3.5\% Br sample showing the TOF experiment detector coverage. Thick red lines show locations of triple axis scans projected onto (H,L) plane.
}
\label{strucrec}
\end{figure}

Disordered quantum magnets are expected to exhibit qualitatively different behaviour from the pure systems.
Apart from straightforward chemical substitution of the magnetic ions for non-magnetic counterparts a more recently implemented idea focuses on creating disorder on peripheral sites involved in superexchange interactions.\cite{Hong10,Wulf2011,Huvonen2012,Yamada11,Yu2011}
Theoretical predictions and scarce experimental work dealing with stability of spin gap\cite{Hyman1996}, lifetime of excitations,\cite{Regnault1995,Xu2007} appearance of in-gap localized states\cite{Sorensen1995,Wang1996} and novel quantum phases\cite{Fisher89} in the presence of disorder necessitate further experiments.
In current paper we discuss the impact of random bond disorder on spin dynamics in a quasi-two-dimensional gapped quantum magnet as seen by inelastic neutron scattering experiments.
Our choice of parent compound is a well-characterized Heisenberg quantum magnet piperazinium hexachlorodicuprate (PHCC). \cite{daoud86,Stone2001,Stone2007} 
It features a layered structure with complex spin network of S=1/2 Cu$^{2+}$ ions bridged by Cu-Cl-Cl-Cu superexchange pathways, see Fig.\ref{strucrec}.
The ground state is a nonmagnetic spin liquid and  
the low energy excitation spectrum consists of a sharp magnon mode with a gap $\Delta$ = 1 meV and a bandwidth of about 1.7 meV. 
Considerable attention has been devoted to magnon lifetimes in PHCC due to observed single magnon decay into two magnons as the two energy scales meet.\cite{Stone2006,Kolezhuk2006,Zhitomirsky2012}
Our focus is on determining the magnon spectrum, it's temperature dependence  and magnon lifetimes in bond disordered PHCC derivates.
We find that although no drastic modification of spectrum occurs bond disorder increases the spin gap, reduces the magnon bandwidth and lifetime.
A random phase approximation treatment of an effective dimer model can account for temperature induced blueshift of the gap and drop of intensity in this complex material.
The most notable effect of bond disorder is increased magnon lifetimes which can be correlated with the single-magnon density of states function.

\section{Samples and experimental methods}
PHCC crystallizes in the triclinic P$\bar{1}$ group with lattice parameters $a=7.9691(5)$\AA, $b=7.0348(5)$\AA, $c=6.0836(4)$\AA\,  and angles 
$\alpha=111.083(3)^\circ, \beta=99.947(3)^\circ, \gamma=81.287(4)^\circ$.
Bond disordered PHCC with 3.5\% and 7.5\% nominal Br content (referred to as PHCX) were prepared by adjusting the HCl:HBr solvent ratio in the starting solutions.
Large fully deuterated single crystals were grown by thermal gradient method.\cite{Yankova2012}
Samples used in inelastic neutron scattering experiments were co-aligned mosaics of 3 samples to within half a degree with total masses of 3.21\,g and 4.27\,g for 3.5\% and 7.5\% Br samples respectively.
Lattice parameters at 1.6\,K were $a=7.86$\AA, $b=6.96$\AA, $c=6.08$\AA  and angles $ \alpha=111.23^\circ, \beta=99.95^\circ,
\gamma=81.26^\circ$ for 3.5\% Br sample and $a=7.90$\AA, $b=6.97$\AA, $c=6.03$\AA, and angles $ \alpha=111.15^\circ, \beta=99.91^\circ,
\gamma=81.29^\circ$ for 7.5\% Br sample.
Time-of-flight neutron scattering experiments were carried out on the direct scattering geometry CNCS instrument\cite{Ehlers2011} at Spallation Neutron Source in Oak Ridge National Laboratory with 4.2 meV incident energy neutrons.
Triple axis neutron spectroscopy measurements were performed on TASP instrument at PSI.\cite{Semadeni2001}
At TASP two modes of operation were used which we denote as the high resolution (hi-res) and the low resolution (lo-res) mode.
In lo-res mode fixed 5\,meV final energy and detector distance of 75\,cm with slit width of 30\,mm was used.
In hi-res mode fixed 3.5\,meV final energy was used and detector analyzer distance was increased to 90\,cm along with narrowing the slit to 20\,mm in front of the detector.
Collimator of 80' was placed after the monochromator and liquid nitrogen cooled Be filter was employed between the sample and analyzer at all times. 
The resolutions determined from incoherent elastic scattering line widths at half maximum were 0.10\,meV in hi-res mode and 0.25\,meV in lo-res mode.
Sample environment was a standard Orange cryostat in all experiments.

\section{Results}
\begin{figure}[!h]
\includegraphics[width=8.5cm]{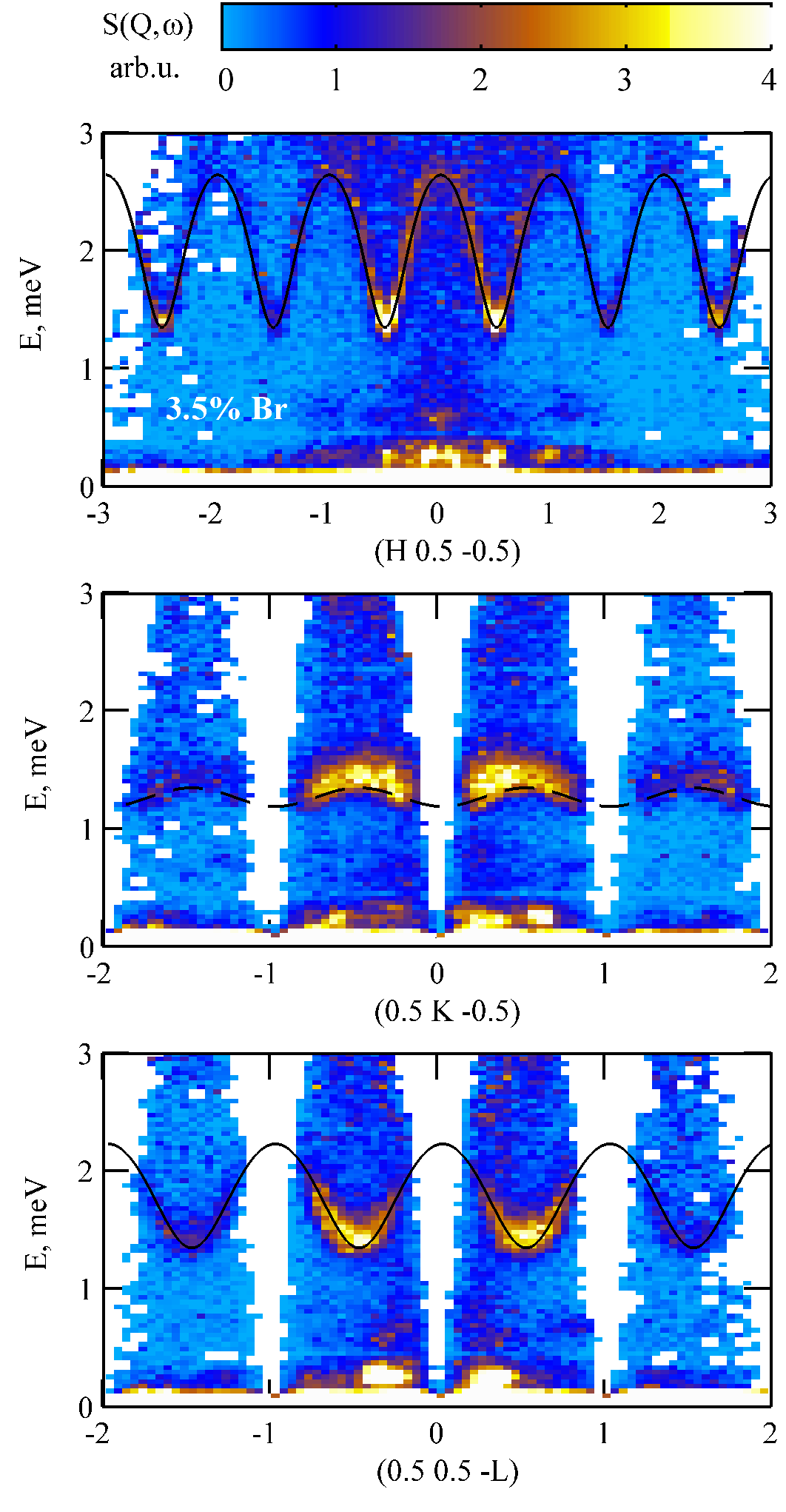}
\caption{(color online) False color intensity maps of the dynamic structure factor $S(\mathbf{Q},\omega)$ for 3.5\%Br PHCX sample along three main reciprocal lattice directions. 
In the above figures the constant reciprocal coordinates denote center points of integration ranges over $\pm 0.1$ r.l.u. Data from distant unit cells has been back-folded to the first Brillouin zone.
Note that some spurious scattering from instrumental origin is visible at wave vectors close to direct beam.
The solid lines are the fitted dispersion curves from triple axis data, Eq. \ref{dispeq} and Table. \ref{disppars}.
The dashed line in the middle panel shows estimated interlayer bandwidth of 0.3\,meV.
}
\label{tof3p5}
\end{figure}

\begin{figure}[!h]
\includegraphics[width=8.5cm]{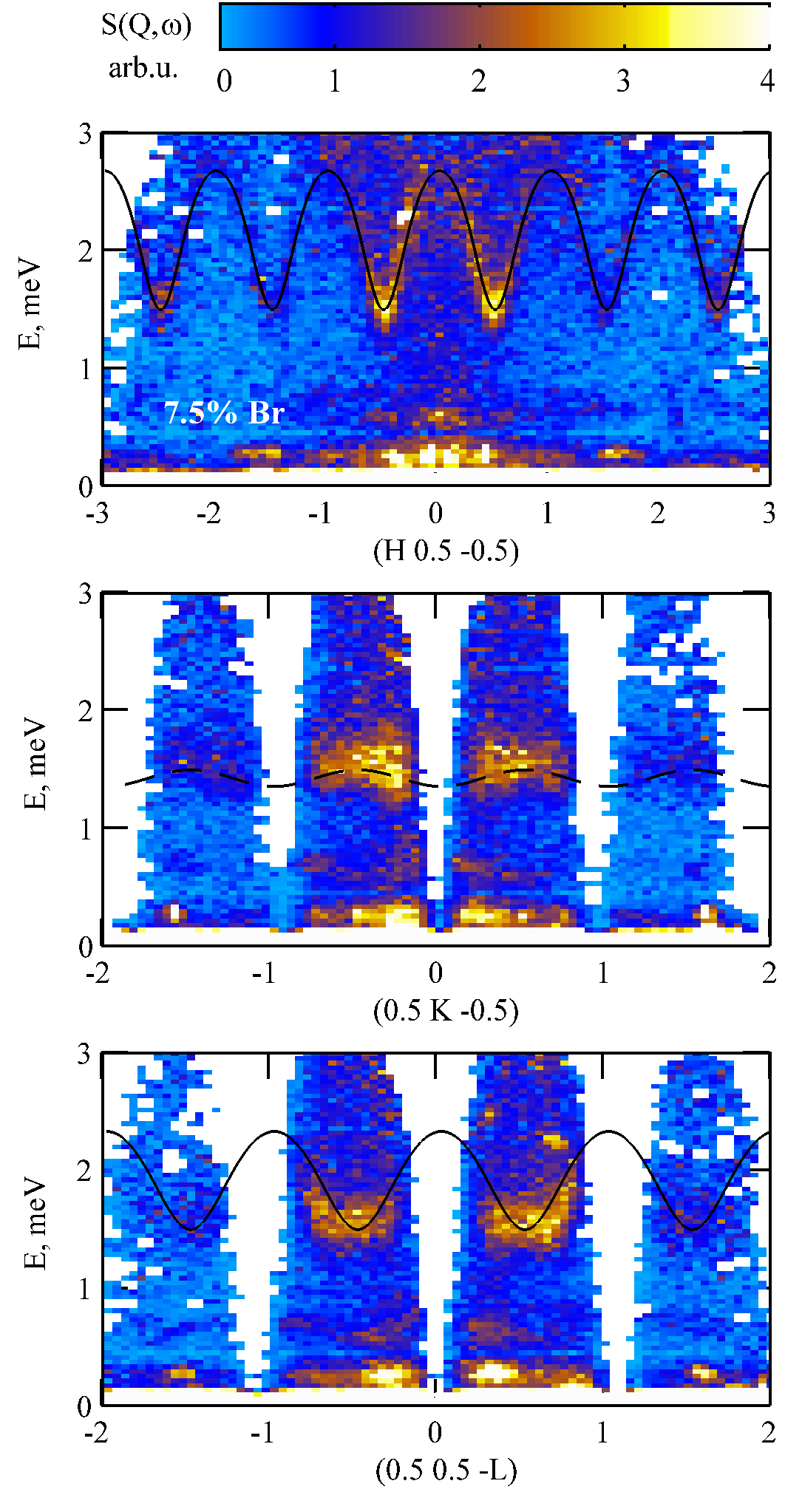}
\caption{(color online) False color intensity maps of the dynamic structure factor $S(\mathbf{Q},\omega)$  for 7.5\%Br PHCX sample along three main reciprocal lattice directions. Coordinates and lines as in Fig. \ref{tof3p5}}
\label{tof7p5}
\end{figure}

Low energy excitation spectrum up to 3\,meV  was recorded by making full 360 degree rotations with 1$^\circ$ step at CNCS for 3.5\% and 7.5\% Br samples. 
Using scattering plane determined by (1,0,0) and (0,1,$\bar{1}$) reciprocal lattice vectors the 190$^\circ$ horizontal and 32$^\circ$ vertical acceptance angles of the detector bank allowed probing of several Brillouin zones simultaneously, Fig.\ref{strucrec}. 
Recorded events were projected to reciprocal sample coordinate system and binned to four-dimensional (H,K,L,E) matrix of size 101x101x101x96 spanning from (-3.5,-2.5,-2.5,-0.5) to (3.5,2.5,2.5,3.3) using Mantid\cite{mantid} program.
Two-dimensional projections from the full spectrum along the three main reciprocal directions are shown in Figures \ref{tof3p5} and \ref{tof7p5}.
In these figures the $S(\mathbf{Q},\omega)$ along the constant reciprocal coordinates was integrated over $\pm 0.1$ range and back-folded to first Brillouin zone. 
The dispersive gapped magnon mode is clearly visible in both samples.
It is evident that no drastic modification to the excitation spectrum occurs in the bond disordered samples.
Note that in the chosen sample orientation and instrument configuration we were unable to probe $K=-L\approx 0$ and equivalent parts spectrum for E$\geq$0.3\,meV.

Magnon linewidth and its wave vector dependence were studied in greater detail by triple axis spectroscopy in conjunction with well established resolution calculation methods using Popovici approximations\cite{Popovici75} embedded in the Reslib package.
Constant-Q scans at the zone center measured with hi-res mode of TASP at the temperature of 1.6\,K were already shown in Fig. 4 of Ref.\onlinecite{Huvonen2012}.
As noted previously the intrinsic magnon linewidth and resonance energy are considerably increased with bond disorder.
Although it is known from recent ESR studies\cite{Glazkov2012,Nafraditobe} that the magnon band is split by 0.038 meV due to anisotropy, this splitting is small enough compared with our experimental resolution to justify using single mode approximation (SMA). 
Morover, it has been established that for the 10\%Br sample the splitting is reduced by a factor of two.\cite{Glazkovprivate}
Hence the broadening is a lifetime effect induced by disorder.

Temperature evolution of the magnon at zone center was studied on TASP using lo-res mode for the broader resonance of the 7.5\% Br sample and in hi-res mode for the 3.5\% Br sample. 
Constant-Q scans across magnon energy at different temperatures are shown in Fig. \ref{tdepspec}.
Blue shift and damping with increasing temperature are evident for both samples.

\begin{figure}[h!]
\includegraphics[width=\columnwidth]{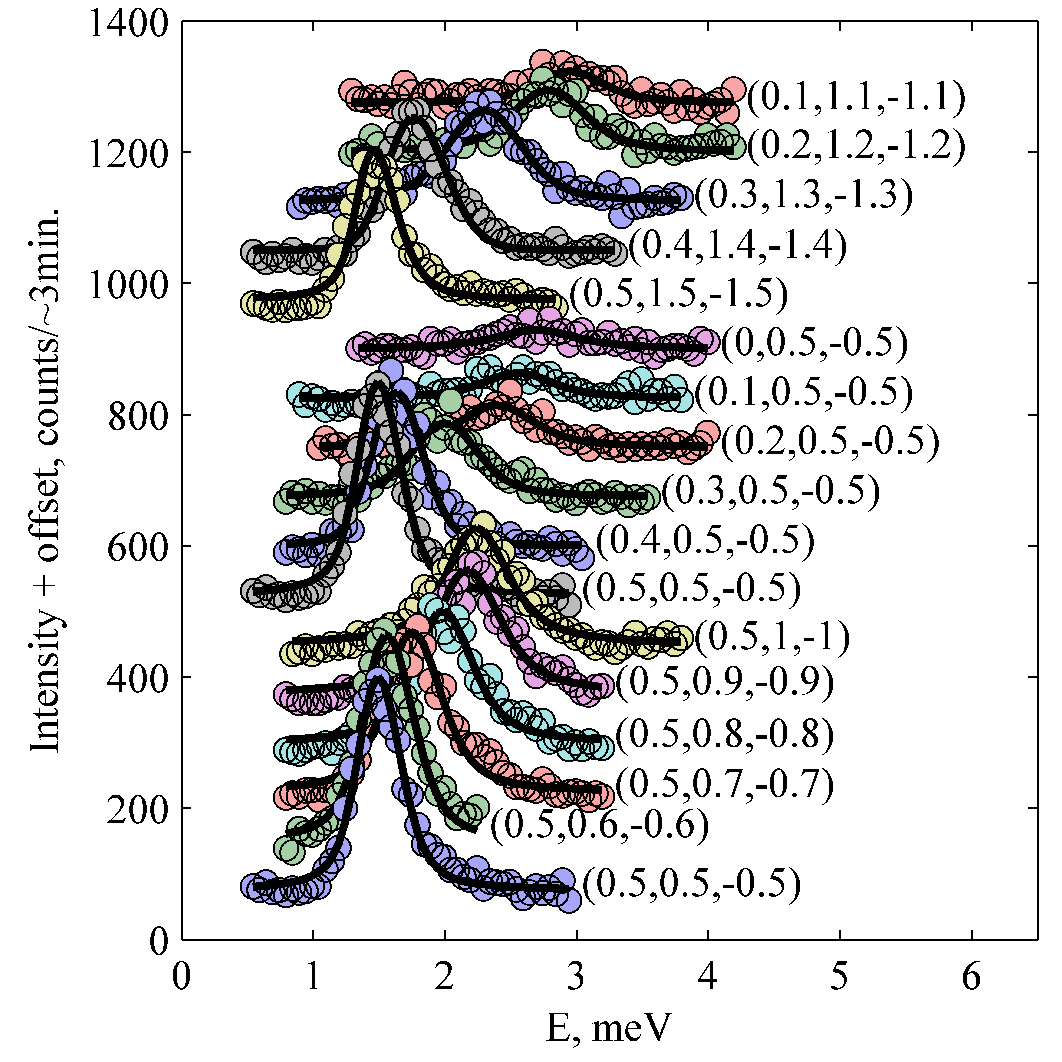}
\caption{(color online) Constant-Q scans across the magnon energy at different wave vectors (shown in parentheses) for the 7.5\% Br sample. Corresponding solid lines are the fitted Lorentzian energy
profiles convoluted with the four-dimensional instrumental resolution
function.}
\label{taspfits75disp}
\end{figure}

\begin{figure}[tb]
\includegraphics[width=\columnwidth]{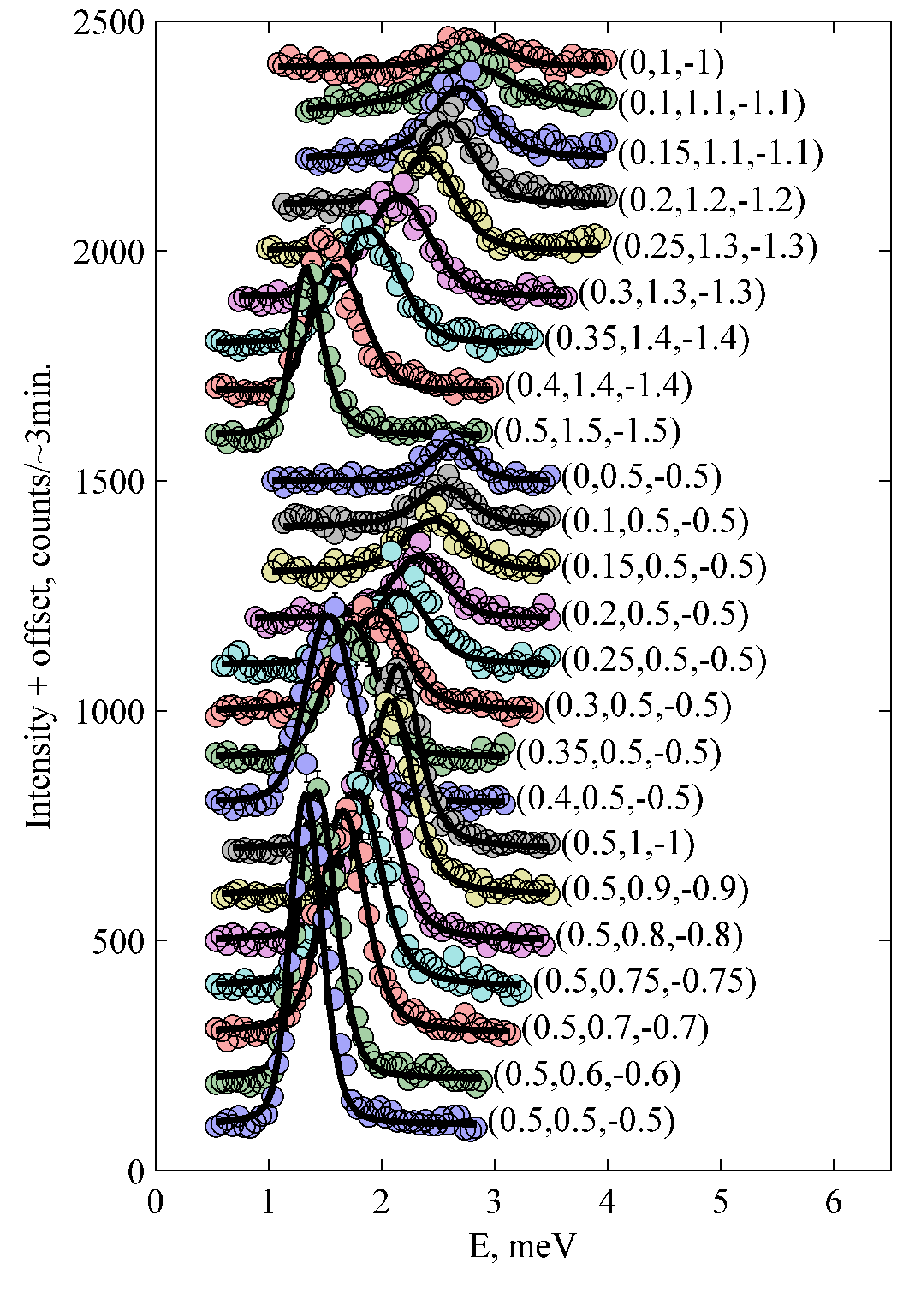}
\caption{(color online) Constant-Q scans across the magnon energy at different wave vectors (shown in parentheses) for the 3.5\%Br sample. Corresponding solid lines are the fitted Lorentzian energy
profiles convoluted with the four-dimensional instrumental resolution
function.}
\label{taspfits35disp}
\end{figure}

\begin{figure}[tb]
\includegraphics[width=\columnwidth]{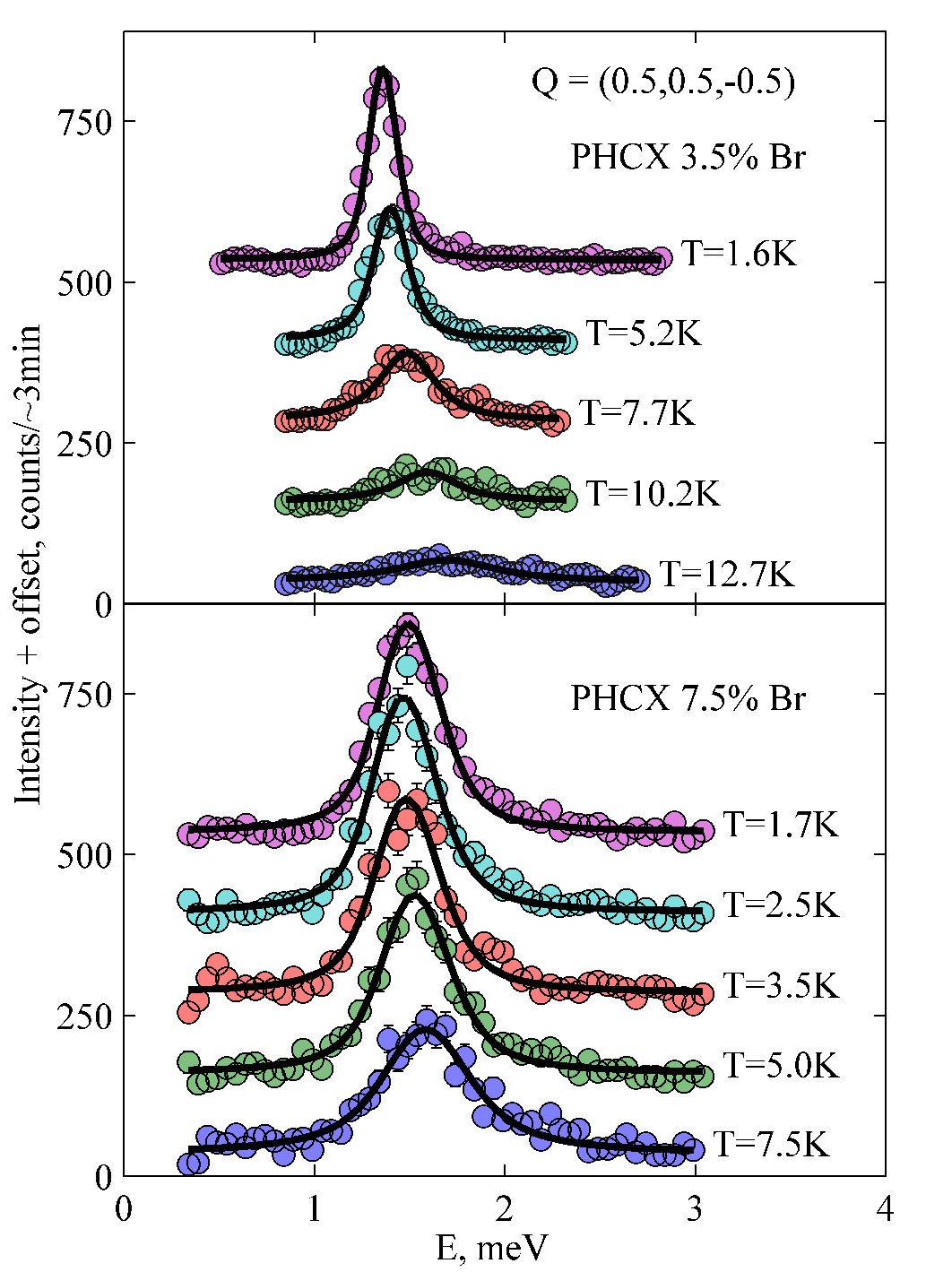}
\caption{(color online) Constant-Q scans across the magnon energy at different temperatures for the 3.5\% and 7.5\% Br samples. Corresponding solid lines are the fitted Lorentzian energy profiles convoluted with the four-dimensional instrumental resolution function.}
\label{tdepspec}
\end{figure}

\begin{figure}[tb]
\includegraphics[width=9cm]{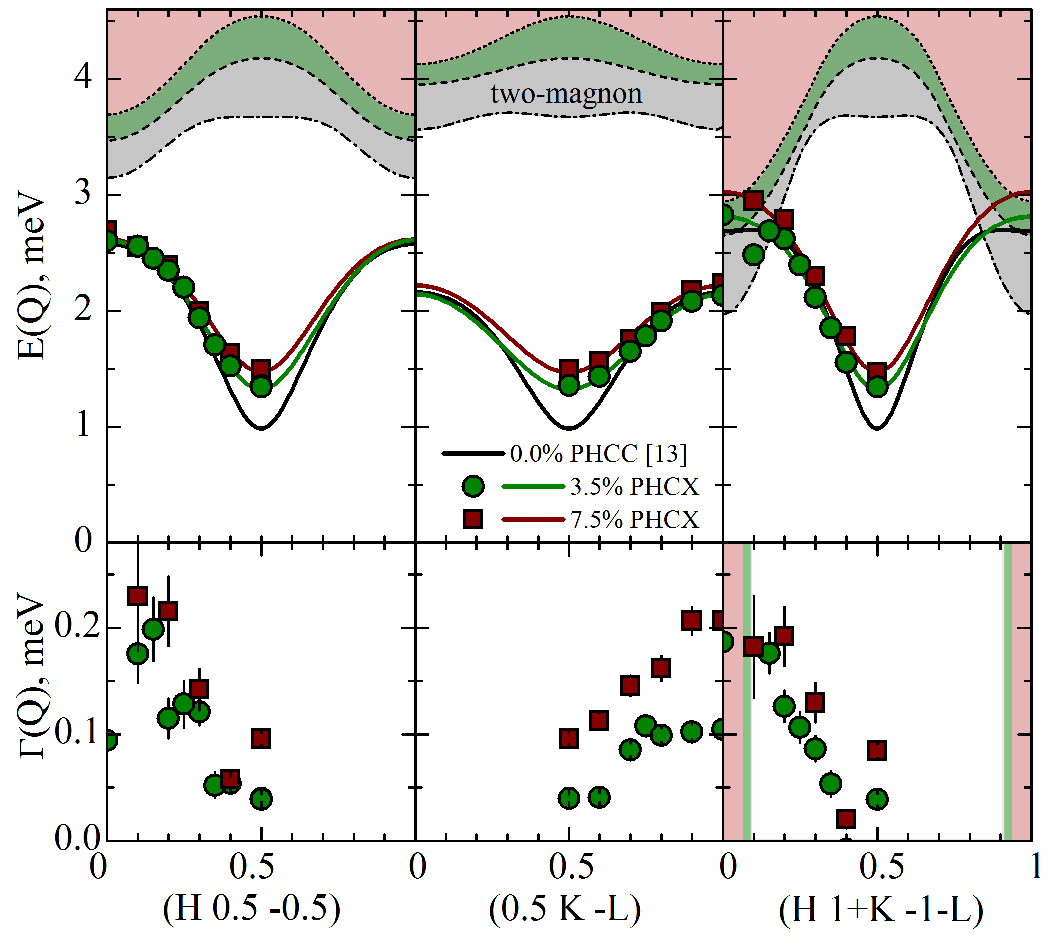}
\caption{(color online) Magnon excitation energies at 1.6\,K and corresponding linewidths are shown with green circles and red squares for 3.5\% and 7.5\% Br samples respectively along three reciprocal space directions. Solid lines are the fitted dispersion curves described by Eq. \ref{dispeq} with paremeters listed in Table \ref{disppars}. Shaded areas bounded by dotted, dashed and dash-dotted lines represent the two-magnon continua for 7.5\%, 3.5\% and 0\% Br samples, respectively.}
\label{disp}
\end{figure}

Triple axis spectroscopy was used to study the magnon wave vector dependence in lo-res mode at T=1.6\,K.
Measured spectra at different wave vectors were iteratively fitted with resolution convoluted Lorentzian lineshapes while adjusting the parameters in dispersion relation, Figs.\ref{taspfits35disp} and \ref{taspfits75disp}.
Following the in-depth analysis of unperturbed PHCC\cite{Stone2001} we parametrized magnon dispersion:
%\begin{equation}
\begin{eqnarray}
&& \hbar\omega(\mathbf{Q}) = (B_0 + \gamma_\mathbf{Q})^\frac{1}{2}, \nonumber\\
&& \gamma_\mathbf{Q} = B_{h}\cos(2\pi h)+ B_{l}\cos(2\pi l)+ B_{hl}[\cos(2\pi(h+l))+ \nonumber\\
&& \cos(2\pi(h-l))]+ B_{2h}\cos(4\pi h)+ B_{2l}\cos(4\pi l)
%+ B_{k}cos(2\pi k)
%\(B_0\)^{\frac{1/2}}$$
\label{dispeq}
%\end{equation}
\end{eqnarray}
%What to do with disp-K???
Best fitting parameters are collected in Table\,\ref{disppars} for comparsion.
Fitted wave vector dependent magnon energies and linewidths are shown as symbols in Fig.\ref{disp}.
The key observations are that for Q=(0.5,0.5,-0.5) the gap increases from 1 meV to 1.36 and 1.50 meV for 3.5 and 7.5\%Br samples respectively.
At the same bandwidth is reduced by about 0.2\,meV from 1.74 meV to 1.53 and 1.58 meV.

The bandwidth reduction is likely a consequence of modification of averaged exchange constants due of induced chemical pressure by Br substitution.
In addition we observed that magnon damping rate increases notably (by approximately a factor of three) when moving towards zone boundaries.
This drastic shortening of magnon lifetime will be discussed in the next section.

Note that there exists small dispersion perpendicular to magnetic layers as seen from TOF spectra, middle panels of Figs. \ref{tof3p5} and \ref{tof7p5}.
However, in our fitting of dispersion relation using triple axis spectroscopy data the used sample orientation did not allow to extract the magnitude of inter-layer coupling independently and hence we fixed it to zero as was done for unperturbed PHCC.\cite{Stone2001}
As the sample orientation was not optimized to measure dispersion along (0,1,0) direction a reliable fit cannot be obtained from TOF spectra either.
From the available data a crude estimate of $\approx$0.3 meV for interlayer bandwidth can be given, shown by dashed lines in the middle panels of Figs. \ref{tof3p5} and \ref{tof7p5}.
The small interlayer coupling renormalizes the exact spin gap and bandwidth values slightly but the conclusions on the disorder effect remain qualitatively intact.

\begin{table}
\begin{tabular}{l|r|r|r} %|d|d}
\hline\hline 

Parameter &	0\% Br [\onlinecite{Stone2001}] &	3.5\% Br &	7.5\% Br \\
in meV$^2$\ &	 &	 &	 \\
\hline 
B$_0$&	5.44(2)&	5.50(4)&	6.12(7) \\
B$_h$&	2.06(3)&	2.17(5)&	2.34(9) \\
B$_l$&	1.07(3)&	1.02(5)&	1.28(8) \\
B$_{hl}$&	-0.39(1)&	-0.20(2)&	-0.05(4) \\
B$_{2h}$&	-0.34(3)&	-0.15(4)&	-0.19(7) \\
B$_{2l}$&	-0.22(2)&	0.00(4)&	-0.04(6) \\
%B$_{k}$&	<0.xx&	<0.1$^*$&	<0.1$^*$ \\

\hline\hline
\end{tabular}
\caption{Best fit parameters to describe the magnon dispersion by Eq.\ref{dispeq} for 3.5\%Br and 7.5\%Br PHCX samples in comparison with PHCC from Ref.\onlinecite{Stone2001} }
\label{disppars}
\end{table}

%\begin{figure}[tb]
%\includegraphics[width=\columnwidth]{ins.png}
%\caption{(Color online) Inelastic neutron scattering intensity at
%(0.5,0.5,−0.5) in the energy transfer range from 0.5 to 2.5 meV
%for x = 3.5\% (green squares) and x = 7.5\% Br (red circles) content
%samples. Corresponding solid lines are the fitted Lorentzian energy
%profiles convoluted with the four-dimensional instrumental resolution
%function. For 0.0\% Br sample the resolution convoluted simulation is
%shown for comparison using parameters from Ref.\onlinecite{Stone2001}.}
%\label{inshires}
%\end{figure}

\section{Discussion}
\begin{figure}[tb]
\includegraphics[width=\columnwidth]{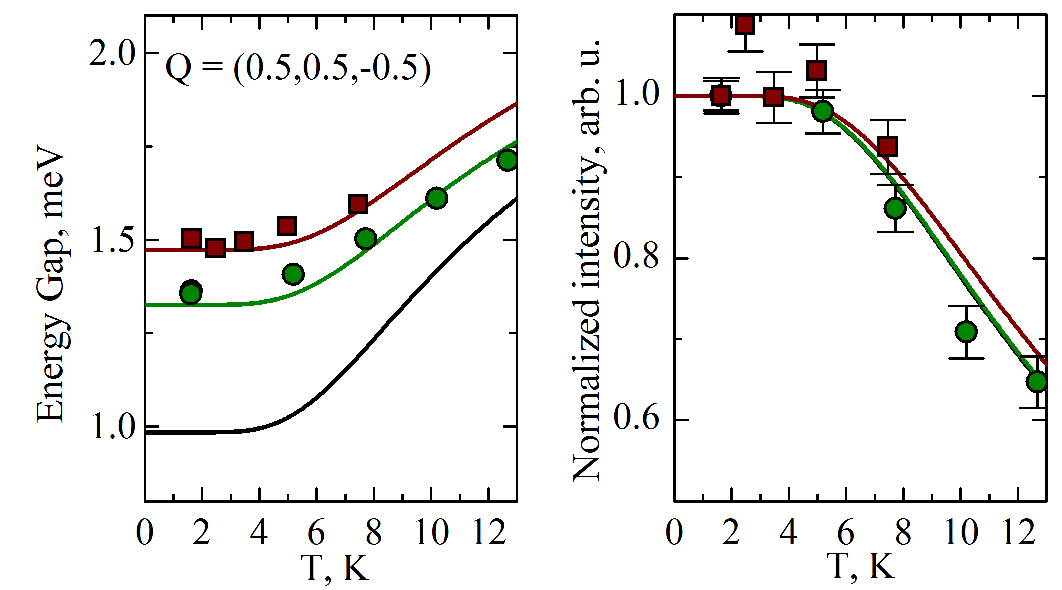}
\caption{(Color online) \emph{Left}: Gap renormalization as a function of temperature: 
%symbols black triangles (Ref.\onlinecite{Nafraditobe}), 
green circles and red squares are the magnon energies at Q=(0.5,0.5,-0.5) for 3.5\% and 7.5\%Br samples. Solid lines are the RPA predictions. \emph{Right}: Intensity of the magnon peak normalized to 1.6\,K intensity. Lines are population differences of singlet and triplet state, see discussion in text. }
\label{rpacalcs}
\end{figure}

Random phase approximation (RPA) has been successful in describing the dispersion of crystal field excitations in several rare-earth compounds,\cite{JensenMackintosh1991} as well as magnon dispersion in several dimerized transition metal based magnets.\cite{Leuenberger1984,Sasago1997,Zheludev1996,Cavadini2000}
Even though the exact microscopic Hamiltonian remains unknown for PHCC and satisfactory coincidence with experimental spectrum has been obtained considering several magnetic Hamiltonians\cite{Mikeska2002,Stone2001} the formal description of the spectrum Eq.\ref{dispeq} is valid for RPA treatment.
Namely, the gap is introduced by the effective dimer coupling $J_0 = \sqrt{B_0}$ and the rest of the spectrum is dominated by smaller magnetic couplings $B_{n(h,l)}$.
Hence we can write for the temperature re-normalized gap\cite{Leuenberger1984}
$\hbar\omega(T)_{\mathbf{Q}} = (B_0 + R(T)\gamma_\mathbf{Q})^\frac{1}{2}$.
Here $R(T)$ is the population difference of singlet ground state and the triply degenerate first excited state $R(T) = n_S - n_T = ({\displaystyle 1-e^{-\beta J_0}})/({\displaystyle 1+3e^{-\beta J_0})}$.
%and $\beta$ standing for inverse temperature.
Intensity of the excitation is directly proportional to the population difference the ground and excited state.
Taking the intensity of the magnon peak at low temperature as scaling constant we expect  $I(T)/I(1.6K) = R(T)$.
The finite temperature renormalization of magnon energy at dispersion minimum the normalized  intensity are shown in Figure \ref{rpacalcs}.
We see that for all samples the temperature induced blue shift and the intensity drop are explained by RPA treatment of an effective dimer model without any adjustable parameters. 
Temperature dependence of magnon linewidths will be discussed in detail elsewhere.\cite{Nafraditobe}

As a result of bond disorder the long lived magnon excitations get notably damped. 
Damping rate is smallest at the zone center Q=(0.5,0.5,-0.5) and increases by about a factor of three at zone boundaries for all measured directions.
This is in contrast to unperturbed PHCC where significant damping occurs at wave vectors where two-magnon continuum has a minimum below single magnon spectrum.\cite{Stonenature2006}
Similar albeit less rapid increase of damping rate has been reported in Haldane spin chain compound CsNiCl$_3$\cite{Zaliznyak2001} as opposed to IPA-CuCl$_3$ where the single quasiparticle spectrum terminates altogether at the wave vector of the crossing.\cite{Masuda2006}
Using the experimentally determined magnon dispersion relation, Eq.\ref{dispeq}, we calculate the energy of the lower boundary for the two-magnon continuum 
$\displaystyle \hbar\omega_{2L}(\mathbf{Q})=\min(\hbar\omega(\mathbf{Q_1})+\hbar\omega(\mathbf{Q_2}))|_{\mathbf{Q_2}=\mathbf{Q}-\mathbf{Q_1}}$
and plot it as an envelopes of the shaded areas in Figure \ref{disp}.
As a direct consequence of bandwidth reduction for PHCX samples the two magnon continuum drops below single particle spectrum only in a very narrow wave vector range, see the shaded areas in bottom right panel of Fig.\ref{disp}.
Hence, we can conclude that two-particle decay cannot account for the observed lifetime reduction in PHCX samples.

\begin{figure}[tb]
\includegraphics[width=\columnwidth]{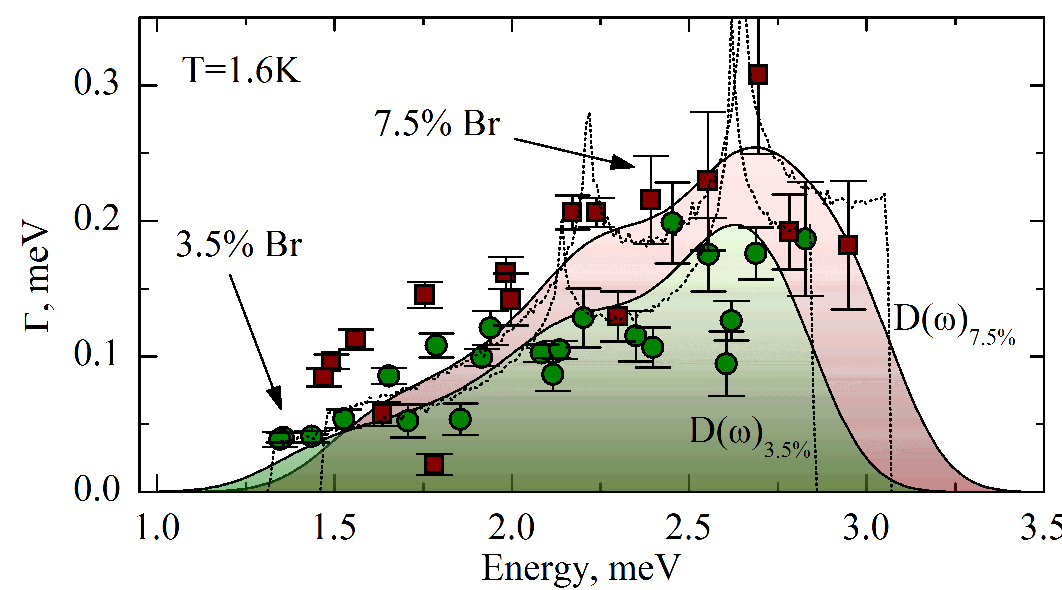}
\caption{(color online) Intrinsic magnon linewidths for 3.5\% Br (green circles) and 7.5\% Br (red squares) PHCX samples collected from Fig.\ref{disp}. Dash-dot lines are the corresponding scaled densities of magnon states. Solid envelopes are the corresponding resolution convoluted densities of state.}
\label{gamdos}
\end{figure}

On the other hand, in bond disordered samples the magnons can scatter off the impurities.
Assuming this process is elastic, the available final state is a magnon with the same energy. 
In this case the scattering rate would follow the density of single magnon states (DOS) at particular energy $\Gamma(\omega) \propto D(\omega)$.
To test this assumption we calculated $D(\omega)$ numerically for both PHCX samples by discretisizing the spectrum described by Eq.\ref{dispeq} and convoluted it with the experimental energy resolution of the inelastic neutron experiment.
These DOS curves are shown with solid line envelopes in Fig.\ref{gamdos}.
The comparison gives remarkably good qualitative agreement.
This result highlights the difference between disorder and temperature induced magnon lifetime shortening.
In a disorder-free system magnon lifetimes shorten due to mutual collisions.
However, in these collisional processes energy of a single magnon needs not be conserved and qualitatively different set of two magnon final states is allowed.\cite{Xu2000,Cavadini2000,Castro2012}

\section{Conclusions}
We have measured the excitation spectrum up to 4 meV for two bond disordered quasi-two-dimensional quantum magnets.
%Induced bond disorder does not have too drastic effects on the excitation spectrum.
No localized states were found to appear within the spin gap and the excitation spectrum remains qualitatively similar to that of the unperturbed material.
However, the spin gap increases and reduction of bandwidth occurs.
We attribute these effects to chemical pressure induced by Br substitution which affects the superexchange strengths.
For both, 3.5 and 7.5\% Br samples RPA explains the temperature induced blue shift magnon mode at the zone centre. 
The drop of magnon peak intensity with temperature is consistent with a singlet-triplet population difference of an effective dimer model.
The long lived quasi-particles of the parent compound acquire finite lifetimes and the damping rate increases away from zone center in all directions.
Damping rate in disordered samples follows the density of single magnon states.

\section{Acknowledgements}
This work is partially supported by the Swiss National Fund under
project 2-77060-11 and through Project 6 of MANEP.
Research at Oak Ridge National
Laboratory's Spallation Neutron Source was supported by the
Scientific User Facilities Division, Office of Basic Energy
Sciences, U. S. Department of Energy.
We thank T. Yankova and Dr. V. Glazkov for their involvement in the early stages of this project.

%\bibliography{danlit,dummypapers,phcc}


\end{document}